\documentclass{appolb}
\usepackage{graphicx}

\usepackage[english]{babel}
\usepackage[utf8]{inputenc}
\usepackage[T1]{fontenc}

\usepackage{amsfonts}
\usepackage{amssymb}

\usepackage{tikz}
\usepackage{nicefrac}
\usepackage{graphicx}
\usetikzlibrary{decorations.pathmorphing} 
\usetikzlibrary{decorations.pathreplacing}
\usetikzlibrary{decorations.markings}
\usetikzlibrary{arrows} 
\usepackage{multirow}

\definecolor{DRed}{RGB}{139,0,0}
\definecolor{DBlue}{RGB}{0,0,139}

\newcommand{\ee}{\,\mathrm{e}}

\begin{document} 

\title{The phase diagram of heavy dense QCD with complex Langevin simulations}  
\author{Gert Aarts, Felipe Attanasio, Benjamin Jäger\thanks{Presented at the Workshop “Excited QCD 2015”, Slovakia, 8-14 March
2015.}
\address{Department of Physics, College of Science, Swansea University,
Swansea, UK}\newline\phantom{stuff}\newline
{Erhard Seiler}
\address{Max-Planck-Institut für Physik (Werner-Heisenberg-Institut), München,
Germany}\newline\phantom{stuff}\newline
Dénes Sexty
\address{Department of Physics, Bergische Universität Wuppertal, Wuppertal,
Germany} \newline\phantom{stuff}\newline
Ion-Olimpiu Stamatescu
\address{Institut für
Theoretische Physik, Universität Heidelberg, Heidelberg, Germany}
}
\maketitle 

\begin{abstract}
The sign problem of QCD prevents standard lattice simulations to determine the 
phase diagram of strong interactions with a finite chemical potential directly.
Complex Langevin simulations provide an alternative method to sample path
integrals with complex weights. We report on our ongoing project to determine
the phase diagram of QCD in the limit of heavy quarks (HDQCD) using Complex
Langevin simulations.
\end{abstract}
  
\section{Introduction}
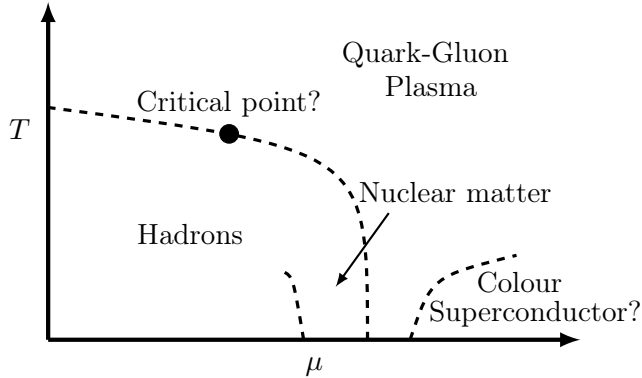
\begin{figure}[!ht]
\centering
\begin{tikzpicture}[scale=1.4]
\begin{scope}[>=latex]
\draw[black,->, ultra thick] (0.0,0.0) -- (0.0,3.20); 
\draw[black,->, ultra thick] (0.0,0.0) -- (5.0,0.00); 
\draw[black,very thick, dashed]  (3.0,0.0) .. controls (3,1.80) and (3,1.80) ..
(0,2.2);  
\draw[black,very thick, dashed]  (2.4,0.0) .. controls (2.3,0.60) and (2.3,0.60)
..
(2.2,0.65); 
\draw[black,very thick, dashed]  (3.4,0.0) .. controls (3.6,0.60) and
(3.6,0.60) ..(4.4,0.8); 
\node[below=0.1cm] at (2.5,0) {$\mu$};
\node[left=0.1cm] at (0,2) {$T$};
\node[left=0.1cm] at (2.0,1.0) {Hadrons};
\node[left=0.1cm] at (4.5,2.7) {Quark-Gluon};
\node[left=0.1cm] at (4.2,2.4) {Plasma};
\node[left=0.1cm] at (4.9,1.4) {Nuclear matter};
\node[left=0.1cm] at (5.0,0.55) {Colour};
\node[left=0.1cm] at (5.75,0.25) {Superconductor?};
\draw[black,->, thick] (3.2,1.2) -- (2.7,0.5); 
\filldraw [black] (1.7,1.95) circle (2.5pt);
\node[above=0.1cm] at (1.7,1.95) {Critical point?};
\end{scope} 
\end{tikzpicture}
\caption{A scenario of the QCD phase diagram.}
\label{Fig1}
\end{figure}

The phase diagram of strong interactions is, despite various efforts, still
largely unknown. Several states of matter are expected to be present, which are
relevant to many phenomena such as the quark gluon plasma, neutron stars and
the evolution of the universe after the big bang. A possible scenario of the
QCD phase diagram is sketched in Figure\,\ref{Fig1}.
A theoretical prediction can guide heavy-ion collision experiments, which might
result in the discovery of different states of matter. However, the
\emph{sign problem} leads to a path integral with a complex weight and thereby
prevents direct determination using standard lattice simulation based on importance sampling.
Complex Langevin simulations, based on stochastic quantisation, might provide a
viable solution to sample path integrals with complex weights~[1 - 16].
In the following we will present an update on our project to determine
the phase diagram of heavy dense QCD (HDQCD), an approximation of QCD in the
limit of heavy quarks, from first principles.

\section{Complex Langevin simulation}

Here, we summarise the basics of the Complex Langevin method, more
details can be found in~\cite{Aarts:2008rr, Aarts:2009dg, Seiler:2012wz,
Aarts:2013uxa}. In analogy to the Hybrid Monte Carlo method, we introduce the
so-called Langevin time $t$, which labels the evolution of observables and degrees of
freedom in this stochastic quantisation. Integrating out the fermion fields
leads to a path integral with a complex weight
\begin{equation}
Z = \int \mathrm{D} U\, \left|\det D\right| \ee^{i
\Theta}\,\ee^{-S_\mathrm{YM}(U)},
\end{equation}
if the chemical potential is real and non zero, since \begin{equation}
\big[\det
D(\mu)\big]^\ast = \det D(-\mu^\ast).
\end{equation}
To incorporate the complex
nature of the path integral in our simulations, we extend the gauge group from
$\mathrm{SU(}3\mathrm{)}$ to $\mathrm{SL(}3,\mathbb{C}\mathrm{)}$. For small
step-sizes $\epsilon$ the gauge links $U_{\mu x}$ are evolved by
\begin{equation}
U_{\mu x}(t+\epsilon) = R(t)\, U_{\mu x}(t), 
\end{equation}
where the update matrix $R(t)$ can be written in term of the Gell-Mann matrices
$\lambda$ and stochastic Gaussian white noise $\eta$
\begin{equation}
R(t) = \mathrm{exp}\left[ i \lambda \left(- \epsilon \, D_{U} S +
\sqrt{\epsilon}\, \eta \right) \right],\label{eqUpdate}
\end{equation}
where the action includes the logarithm of the determinant. Here we study QCD
in the limit of heavy quarks, for which the fermion determinant can be written
in terms of the (conjugate) Polyakov loops $\mathcal{P}_{\vec{x}}$ 
and $\mathcal{P}^{-1}_{\vec{x}}$

\begin{equation}
\det D(\mu) = \prod_{\vec{x}} \det \left( 1 + h\,\ee^{\mu/T} \,
\mathcal{P}_{\vec{x}} \right)^{2}\det \left( 1 + h\,\ee^{-\mu/T}\,
\mathcal{P}^{-1}_{\vec{x}} \right)^{2},
\end{equation} 
with $h = \left( 2 \, \kappa \right)^{N_\tau}$. For the gluonic part of the
action we use the full Wilson gauge action. To avoid runaway trajectories into
the non-compact extension of $\mathrm{SU(}3\mathrm{)}$, we apply adaptive step-size
scaling~\cite{Aarts:2009dg} and adaptive gauge
cooling~\cite{Seiler:2012wz,Aarts:2013uxa}.
Too many large excursions into the imaginary directions have been identified
to cause the Complex Langevin method to fail by converging to incorrect results.
It can be shown, that  if the action is holomorphic and suitably confined in 
the complex extension of $\mathrm{SU(}3\mathrm{)}$, Complex Langevin
simulations are expected to converge to the correct
results~\cite{Aarts:2009uq,Aarts:2011ax}.  The logarithm of the determinant 
causes poles in the derivative of the action, and thereby prevents  the
aforementioned proof to be applied.
Nevertheless, recent work~\cite{Mollgaard:2013qra,
Splittorff:2014zca,Nishimura:2015pba} has shown that especially for large quark
masses this ambiguity will not affect Complex Langevin dynamics. We still
monitor the distributions of the observables and the so-called unitnorm,
\begin{equation} 
\mathrm{unitnorm} = \mathrm{Tr} \left( U U^{\dagger} - \mathbb{I} \right)^2, 
\end{equation} to avoid runaway trajectories in our simulations.

\section{Numerical setup and results}

We study the phase diagram of heavy dense QCD for fixed lattice spacing and
the simulation parameters are given in Table\,\ref{tab}.
\begin{table}[h!]
\begin{center}
\begin{tabular}{|c|c|c|}
    	\hline
    	$\beta = 5.8$   & $N_f=2$ & $V=8^3 \times N_\tau$ \\
    	$\kappa = 0.04$ & $\mu = 0.0 - 3.2$ & $N_\tau = 2-32$\\
    	   	\hline
\end{tabular}
\caption{Summary of simulation parameters.}
\label{tab}
\end{center}
\end{table}
For HDQCD the expected critical chemical potential $\mu_c$ (in lattice units) is
related to the bare quark mass by 
\begin{equation}
\mu_c \sim m_q \equiv - \ln(2 \kappa) = 2.53.
\end{equation} 
We have improved our previous results~\cite{Aarts:2014kja} by considering larger
Langevin trajectories, with a maximum Langevin time of $500$. The interval up to 
$100$ Langevin time has been discarded to to remove thermalisation effects.
Using adaptive step-sizes $\epsilon$ we find typical values of $\epsilon \sim
10^{-4}$. We have determined the observables every $\delta t = 10^{-2}$, which
corresponds to approximately $100$ sweeps in between measurements. Including 
auto-correlation we have at least $2000$ independent configurations for each simulation.
\begin{figure}[!ht]
	\centering
	\vspace{-0.5cm} 
	\begin{minipage}{0.95\linewidth}
	\centering
	\includegraphics[width=\linewidth]{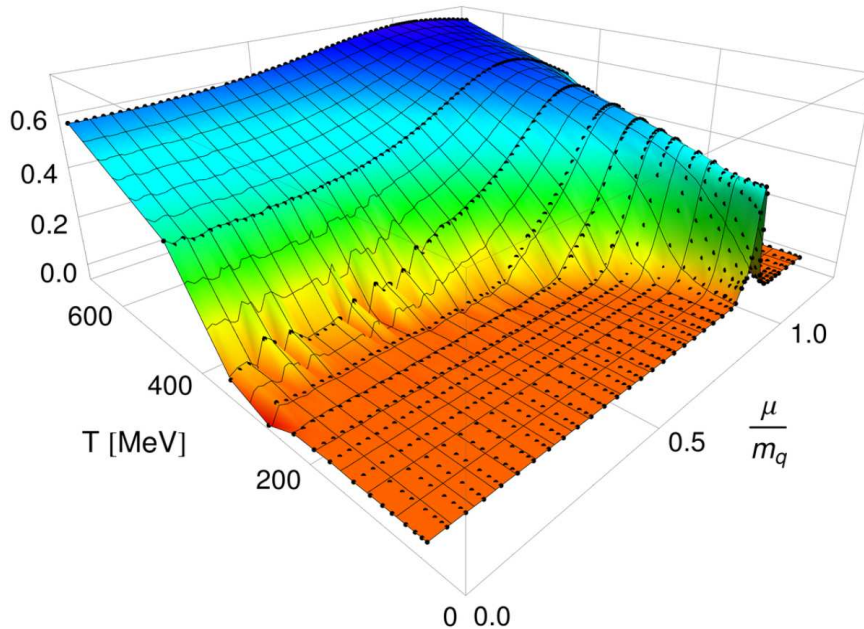}
	\end{minipage}
\caption{The Polyakov loop as function of $T$ and $\mu$.}
\label{Fig3}
\end{figure}
Figure\,2 shows the expectation value of the Polyakov loop as a
function of temperature $T$ and the chemical potential $\mu$. The temperature 
has been converted to physical units using the lattice spacing  of
$a\sim0.15\,\mathrm{fm}$, which has been determined  using the Wilson
flow~\cite{Sexty:2013ica,Borsanyi:2012zs}. Each black  point in Figure\,2 is the
result of a dedicated simulation. The Polyakov loop shows a clear signal for
the deconfinement transition and a transition to higher densities. 
At high densities, $\mu/m_q \geq 1$, the Polyakov loop
drops again. This behaviour is an expected lattice artefact, at which every
lattice site has been filled with the maximum number of fermions allowed by the
Pauli principle. The coloured surface is a cubic interpolation to the
individual simulations. The resolution in temperature is quite limited in the
fixed lattice spacing approach, since the temporal extent is by construction an
integer. \begin{figure}[!ht] 
	\centering 
	\hspace{-0.0cm} 
	\begin{minipage}{0.95\linewidth}
	\centering
	\includegraphics[width=\linewidth]{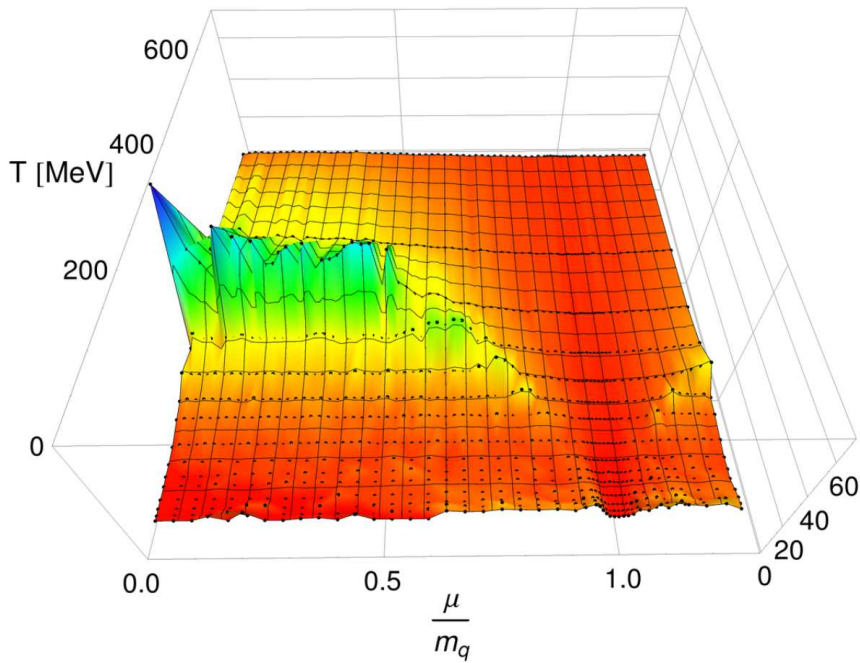}
	\end{minipage}   
\caption{The Polyakov loop susceptibility as function of $T$ and $\mu$.} 
\label{Fig4}
\end{figure}
Figure\,3 shows the equivalent plot for the susceptibility of the
Polyakov loop, which directly maps out the boundary of the phase diagram of HDQCD.
The deconfinement transition appears to be quite broad, which is caused by
our limited resolution for large temperatures and the subsequent interpolation.
We find a clearer signal for the transition to higher densities, which almost
disappears on the plotted scale, using the susceptibility of the fermion density
\begin{equation}
n = \frac{1}{N_\tau N_s^3} \frac{\partial\, \mathrm{ln}\, Z}{\partial \mu}.
\end{equation}

\section{Conclusions and Outlook}

Complex Langevin simulations provide a viable method to determine the phase
diagram of heavy dense QCD from first principles. Further work includes the 
identification of the order of the transitions by varying the simulation
spatial volume and studying the Binder cumulant. Simulation at different lattice
spacing will improve the resolution in the temporal direction and allow to
asses the size of lattice artefacts. The ultimate goal is to repeat these
simulations for fully dynamical QCD~\cite{Sexty:2013ica,Aarts:2015oka} and study
the phase diagram of QCD itself. In perspective of this goal, the work here can
be considered as blueprint for further studies and as proof of principle.
\newline \phantom{stuff}\newline

{\bf Acknowledgments:} 
We are grateful for the computing resources made available by HPC
Wales and by STFC through DiRAC computing facilities. This work is
supported by STFC, the Royal Society, the Wolfson Foundation and the Leverhulme Trust. 
FA is grateful for the support through the Brazilian government program "Science
without Borders" under scholarship number Bex 9463/13-5. BJ acknowledges
financial support from the College of Science Research Fund at Swansea
University.

\end{document}